\documentclass[11pt,twoside]{article}

\usepackage{asp2006}
\usepackage{psfig}
\usepackage{lscape}

\begin{document}

\title{Observational evidence for dust growth in proto-planetary discs}  
\author{Gwendolyn Meeus}   
\affil{AIP, An der Sternwarte 16, D-14482 Potsdam, Germany}  

\begin{abstract}
The dust in the interstellar medium, that provides the material for forming
stars - and circumstellar discs as a natural by-product - is known to have
submicron sizes. As these discs are the sites of planet formation, those small
grains are predicted to grow to larger entities when the stars are still
young.  I will review evidence for the first steps in grain growth in
proto-planetary discs around young stars, based on recent {\it Spitzer} and
ground-based infrared observations. First, I will discuss disc and dust
properties in Herbig Ae/Be stars, and then move to the lower-mass T Tauri
stars and the brown dwarfs. Here, objects of different star-forming regions
are compared, and the influence of the stellar parameters and environment on
dust evolution, as witnessed by the observed dust characteristics, is
discussed.
\end{abstract}

\section{Introduction}
\label{intro}

The research of dust growth in the disc of young stellar objects is motivated
from the observations of dust in the interstellar medium (ISM):
\cite{kemper2004} showed that the crystalline fraction of the ISM dust is
smaller than 2\%, and the size of the amorphous silicates is smaller than 0.1
micron (see Fig.~\ref{f_sizes}, left panel). It is this dust that eventually
will constitute the proto-planetary disc, so that, {\it initially}, the dust
grains in circumstellar (CS) discs must have small sizes too. However, as
planets are expected to form in these discs, the grains will have to grow to
larger sizes, forming planetesimals, and finally planets. This process,
however, is not yet well understood, especially not the timescale over which
it happens, or which physical properties of the star or the environment would
impede or hasten the growth of dust particles. It is those first steps of
grain growth, from submicron sizes to sizes of a few microns, that I will
discuss in this paper.


First, we will give a short introduction into the mineralogy of
astronomical dust, and show how we can retrieve information about 
the dust properties from infrared spectroscopy. In a next step, we 
will discuss the disc and dust properties of the bright Herbig Ae/Be 
stars, and then we will apply the same technique for the solar-mass
T Tauri stars and the substellar mass brown dwarfs. 

\section{Mineralogy in a nutshell}
\label{s_minera}

\begin{figure}
\centerline{\hbox{
\psfig{figure=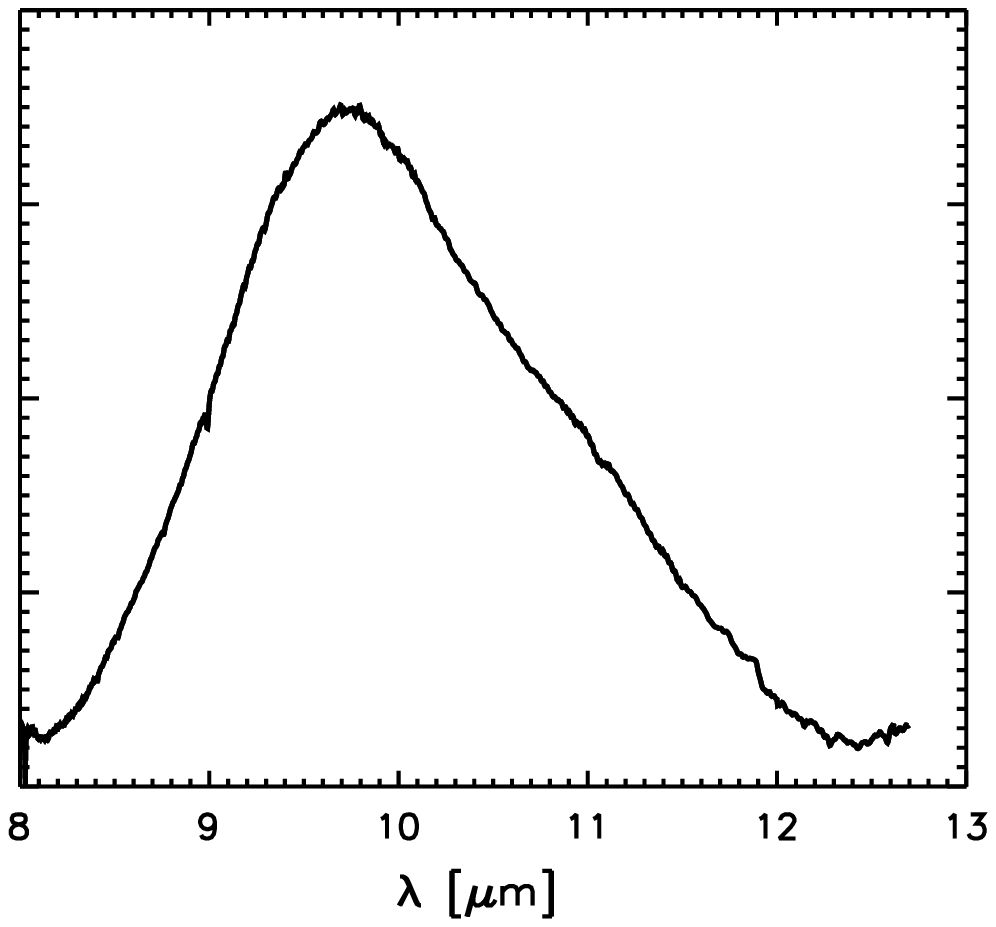,width=5cm,height=5cm}
\psfig{figure=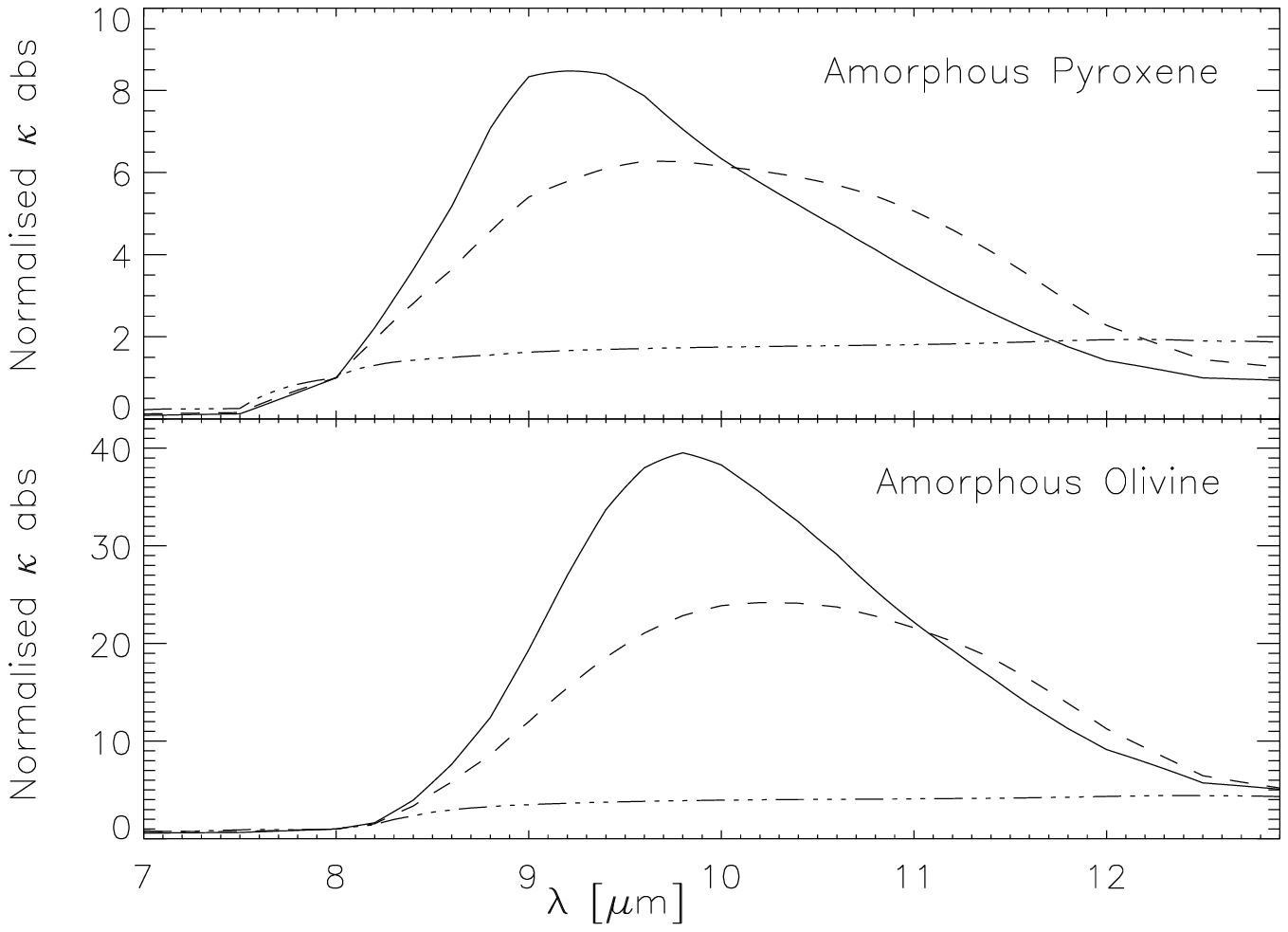,width=7cm}}}
\caption{Left: The optical depth towards the galactic center is clearly 
caused by small, amorphous silicate grains, as indicated by the shape
observed \citep{kemper2004}. 
Right: Absorption coefficients of amorphous silicates with a pyroxene and 
olivine stoichiometry, in 3 different sizes that are relevant in the 10 
micron window \citep{dorschner1995}. Solid
line: 0.1 micron, dashed line: 1.5 micron and dashed-dotted line:
6.0 micron. With increasing size, the feature shifts towards longer
wavelengths with an increasingly important red shoulder, and flattens.}
\label{f_sizes}
\end{figure}

Astronomical dust is either oxygen or carbon rich. Oxygen-rich dust particles
are mainly silicates, which can be either iron or magnesium-rich, have a
crystalline or amorphous structure, and a wide variety in shapes and sizes. In
the 10 micron observational window, the most important dust species such as
pyroxenes, olivines and silica show features that can be used to derive their
size, composition and structure (see Fig.~\ref{f_sizes} for an example of
pyroxene and olivine features). For this purpose, optical constants that are
measured in the lab are of great importance (e.g. \cite{tamanai2006}).

\section{Herbig Ae/Be Stars}
\label{s_haebes}

\begin{figure}
\centerline{\psfig{figure=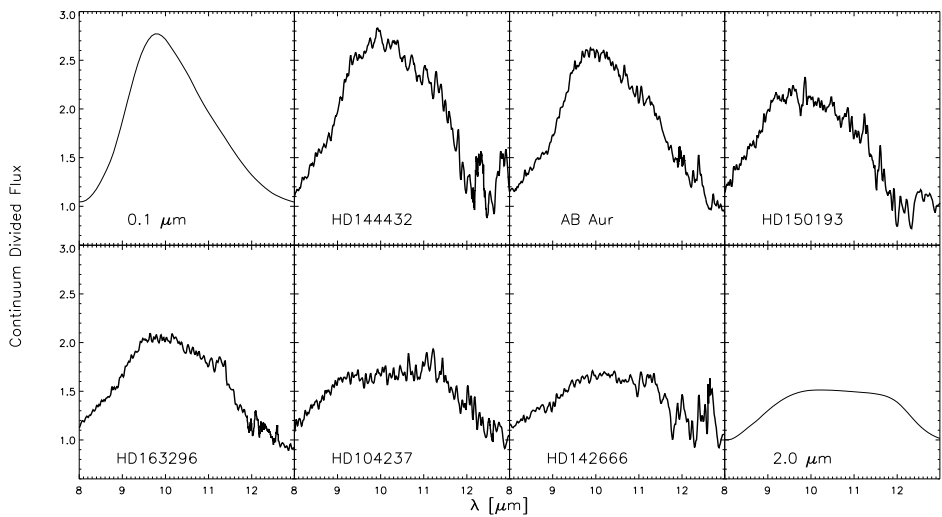,width=13cm}}
\caption{The 10 micron feature for a sample of Herbig Ae/Be stars. They show
a wide variety in shape and strength, pointing to the presence of both small
and large grains in their discs, with a varying amount of crystallinity
\citep{meeus2001}.}
\label{f_haebefeatures}
\end{figure}

Infrared spectroscopy knew a large step forward through the launch of the
Infrared Space Observatory (ISO). For the first time, the dust in
circumstellar discs could be studied in detail. In particular, the discs of
the bright Herbig Ae/Be stars, a class of pre-main sequence stars with
spectral type A or B that are marked by the presence of the H$\alpha$ emission
line and an IR excess due to dust, were studied extensively between 2 to 45
micron. The dust particles that emit in this wavelength region are warm (a few
100 K), and are located in the optically thin disc atmosphere.  The bulk of
the disc material, however, is cold (below 100 K) and resides in the midplane
of the disc, which is optically thick, hence invisible to us. 

ISO unveiled a wide variety of dust features in Herbig Ae/Be stars: different
shapes and strengths, pointing to a wide range in grain size and crystallinity
were observed \citep{meeus2001,bouwman2001}. In Fig.~\ref{f_haebefeatures}, we
give a few examples of the emission features observed in the 10 micron window.
\cite{boekel2003} further related the 10 micron feature strength and shape
(triangular versus a more flattened shape) at 10 micron to the grain size of
the silicates, with the aid of laboratory spectra, and found clear evidence
for grain growth in HAEBEs.

\begin{figure}
\centerline{\psfig{figure=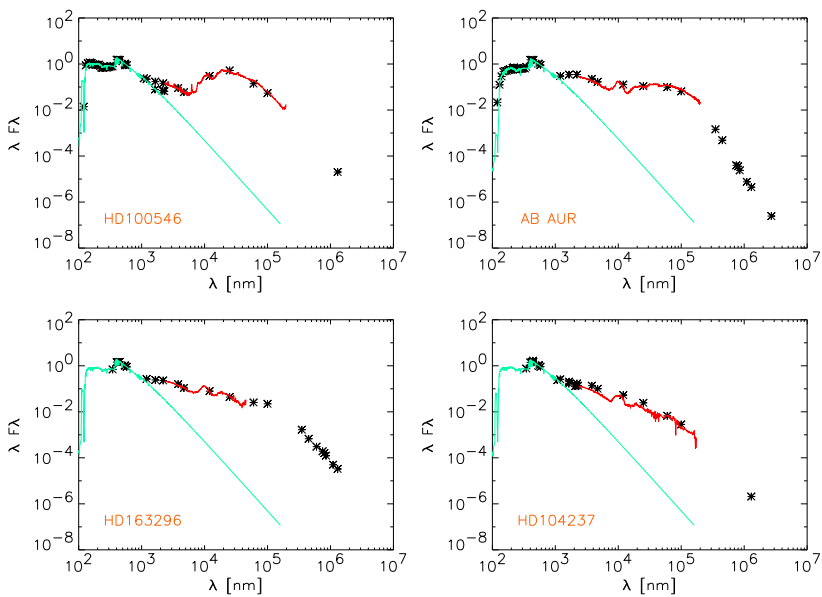,width=11cm,angle=90}}
\caption{The spectral energy distributions of selected 
Herbig Ae/Be stars. The slope in the millimetre region 
is much steeper for AB Aur than for HD163296, pointing
to larger cold grains for the latter star \citep{meeus2001}.}
\label{f_haebesed}
\end{figure}

Radiative transfer models of the disc structure and evolution predict that
when small grains, located in the disc atmosphere, grow to larger particles,
they will gradually settle towards the disc midplane.  This will be
accompanied by a flattening of the (initially) flared disc
\citep{dullemond2004}. This process has been confirmed through observations:
objects which have a smaller mid-IR excess (pointing to a less flaring disc)
on average have a shallower slope at millimetre wavelengths, indicating larger
sizes for the cold grains \citep{acke2004}.

\section{T Tauri Stars}
\label{s_tts}

T Tauri stars are the lower-mass counterparts of the Herbig Ae/Be
stars, with spectral types between G and M, thus temperatures below
$\sim$ 6000 K. Recent {\it Spitzer} observations allowed for a
characterisation of a large sample of T Tauri stars, thanks to its
good sensitivity. It soon became clear that T Tauri stars have similar 
dust properties as the Herbig Ae/Be stars, although they have much 
smaller luminosities: some objects show nearly unprocessed dust, while 
others show larger grains and a substantial amount of crystalline silicates 
\citep{meeus2003,kessler2006}.

\begin{figure}
\psfig{figure=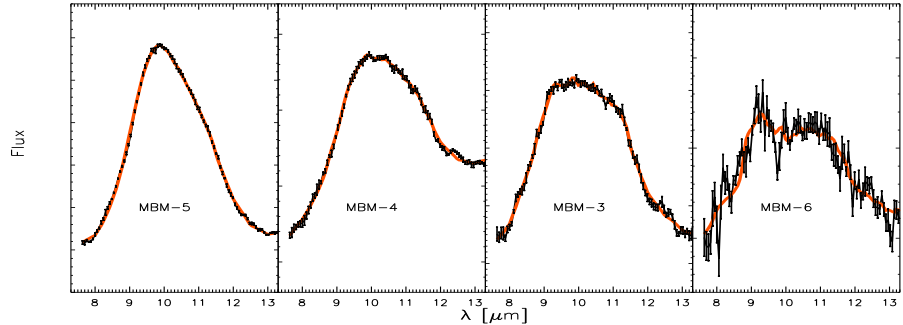,width=12.5cm,height=4.5cm,angle=90}
\caption{The 10 micron feature for four T Tauri stars in MBM12. Some
features show evidence of small, amorphous grains, while others are
obviously caused by larger and more crystalline dust.  
We also overplot a fit of the feature by the TLTD method.}
\label{f_mbm12_sili}
\end{figure}

\begin{figure}
\centerline{\psfig{figure=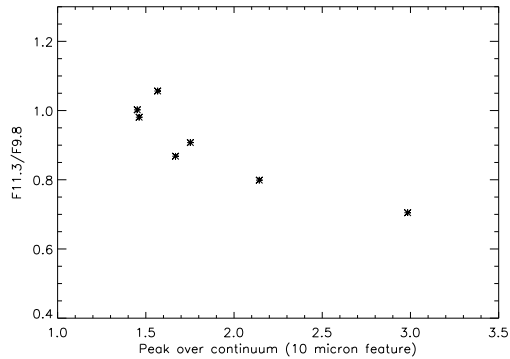,width=7cm}}
\caption{The shape of the 10 micron feature, determined by the flux ratio 
at 11.3 and 9.8 micron, in function of the feature strength, determined 
by the peak over continuum ratio for T Tauri stars.}
\label{f_peakco}
\end{figure}

\begin{figure}
\centerline{\hbox{
\psfig{figure=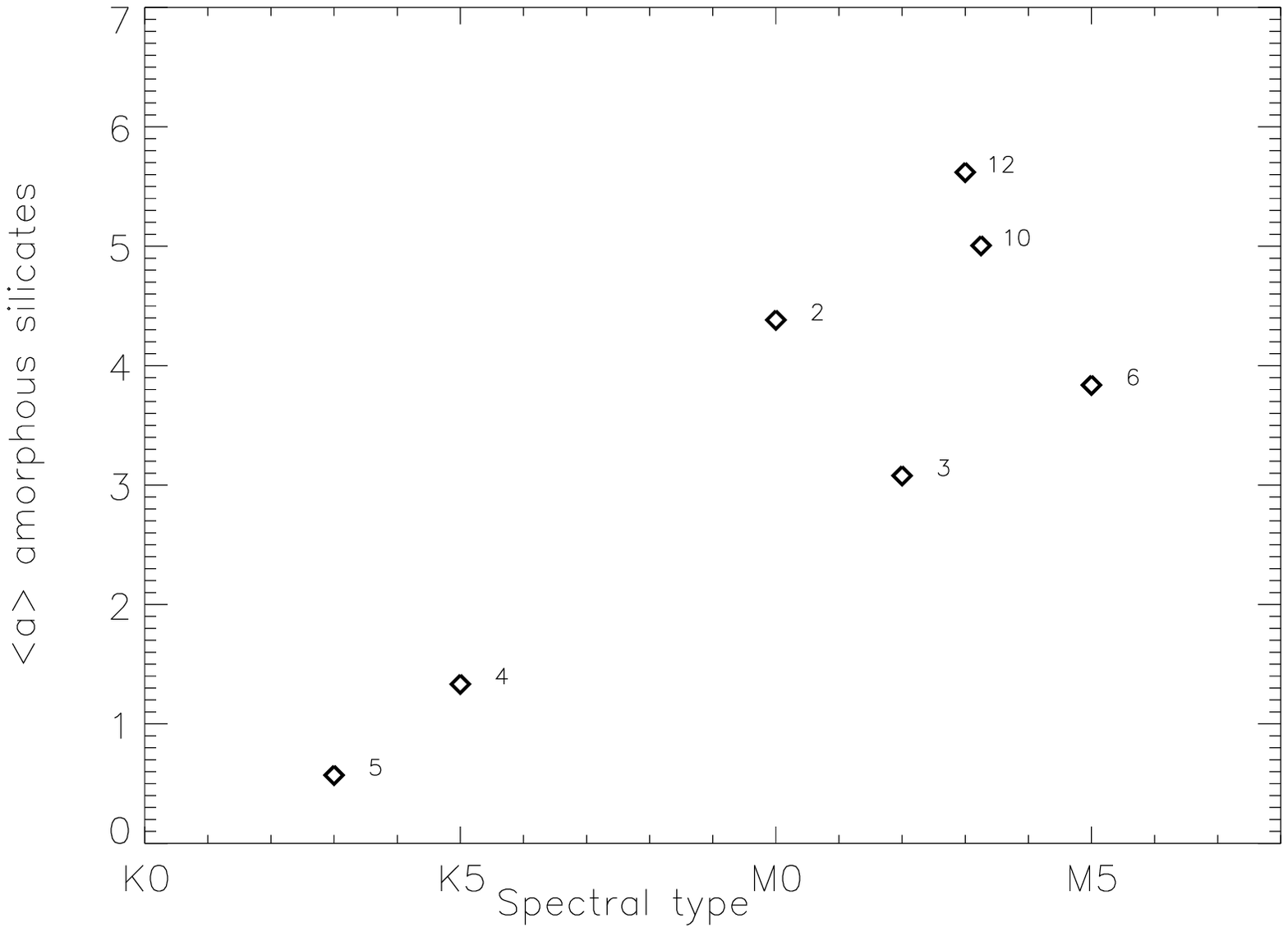,height=5cm}
\psfig{figure=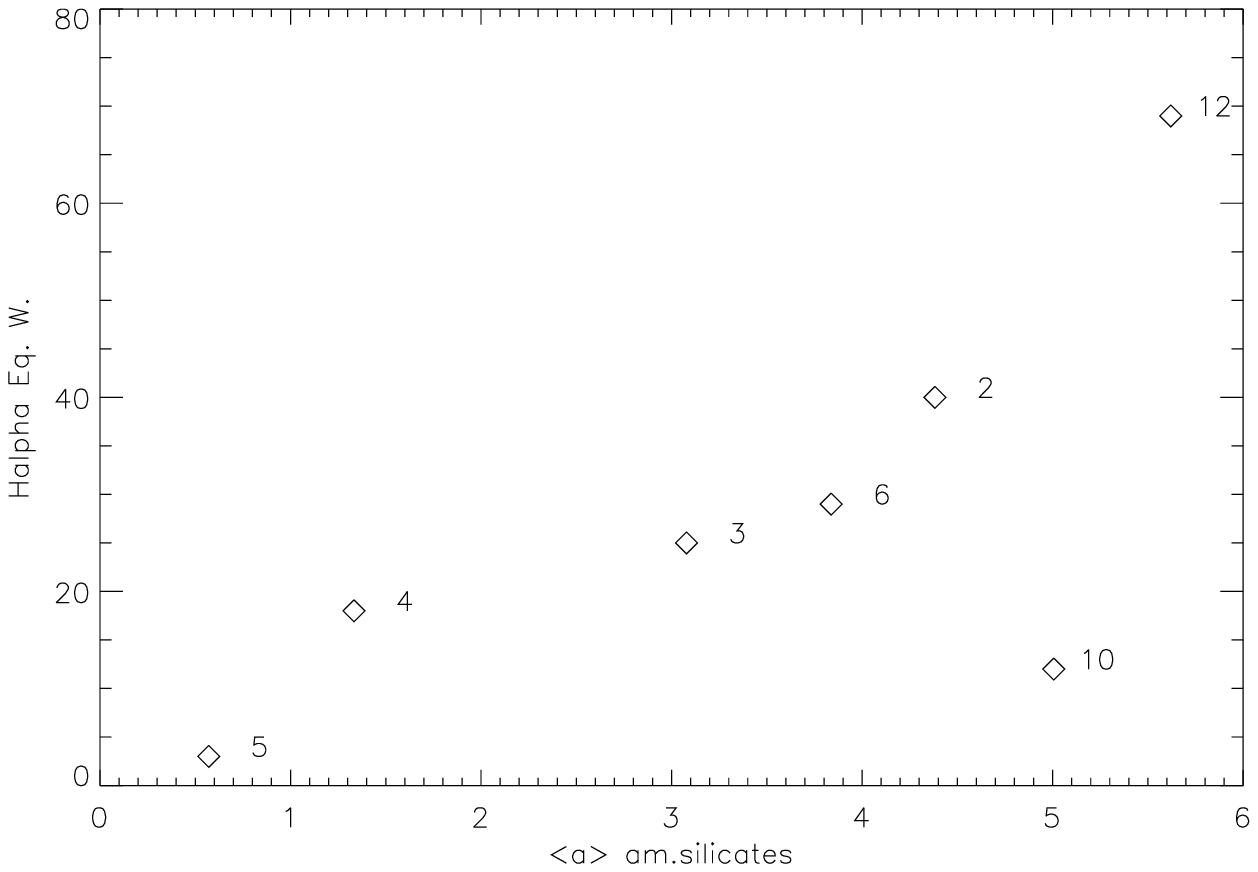,height=4.9cm}}}
\caption{The size of the amorphous silicates, as determined
by the TLTD fit, in function of the spectral type (left) and the H$\alpha$ 
equivalent width (right). Although the sample is really small, we do see a 
trend between both variables: the later the spectral type, the larger the
grains observed, and the stronger the H$\alpha$ line, the larger the 
grain sizes.}
\label{f_amsize}
\end{figure}

In an unbiased sample of 12 TTS in the star forming region MBM12, with an age
of 2~Myr, dust processing was also observed in different stages
\citep{meeus2009}. In Fig.~\ref{f_peakco}, we show the relation between the
shape and the strength of the 10 micron feature, indicating grain grows in
these TTS discs.  We derived the properties of the dust causing the 10 micron
feature by modelling the feature, using the two layer temperature distribution
method (TLTD) by \cite{juhasz2008} and including the following dust species in
the fit: amorphous silicates, forsterite, enstatite and silica, in sizes of
0.1, 1.5 and 6.0 micron. We found that those objects that have the latest
spectral types (lowest temperatures), tend to have the largest grains (see
Fig.~\ref{f_amsize}), a relation that was first remarked by
\cite{kessler2006}. This is merely because, when observing at 10 micron -
tracing dust grains with temperatures of a few hundred Kelvin - we trace a
region that is much closer in for those objects which have a lower luminosity
than for those with a higher luminosity. For instance, for T Tauri stars, this
is of the order 0.5 AU, while for Herbig Ae stars, radial distances of a few
AU are reached. When one, furthermore, considers that the density in the disc
decreases with radial distance from the central star, and that grain growth
increases with density, then it is natural to expect that the lower luminosity
sources show more grain growth, when observing the 10 micron region
\citep{kessler2007}.

In MBM12, the degree of flaring, as derived from the flux ratio at 24 and 8
micron is found to relate to the grain size, as derived from the 10 micron
feature \citep{meeus2009}.  Furthermore, those sources that are most
turbulent and accreting, as indicated by the equivalent width of the H$\alpha$
feature, are found to host the largest grains in their disc atmosphere, while
the lowest-accreting source have more ISM-like silicates (see
Fig.~\ref{f_amsize}), as was already noted by \cite{sicilia2007} in another
sample of T Tauri stars.

\section{Brown Dwarfs}
\label{s_bd}

Even for the very faint brown dwarfs, {\it Spitzer} provided spectra, but only
for small samples. In Chamaeleon I, with an age of 2 Myr, 6 out of 8 brown
dwarfs show an IR excess, but the amount of flaring is quite low for most of
them.  Both grain growth and crystallisation is observed to occur routinely in
this sample \citep{apai2005}, which might be attributed to their low
luminosity: in the innermost regions of their discs, where the 10 micron
radiation originates for brown dwarfs, it is likely that most of the dust has
grown to larger sizes. However, the large crystallinity observed could also
(partly) be a contrast effect, as larger grains cause weaker spectral
features, so that the crystalline features become more clear. Only at an age
of a few Myrs older, the discs and dust around brown dwarfs appear to have
evolved drastically: in the star forming region Upper Scorpius, with an age of
5 Myr, most discs are flat, pointing to a large degree of dust settling, while
the 10 micron feature is mainly absent or, in the rare case it is present,
very weak \citep{scholz2007}. At an age of 10 Myr, the brown dwarfs do not
even show the feature anymore \citep{morrow2008}. These observations show that
dust and disc evolution around brown dwarfs occurs on a much faster timescale
than in T Tauri stars or Herbig Ae/Be stars.

\section{Conclusions}
\label{s_conc}

When discussing disc and dust evolution, parameters that come to mind as
likely important players are age and effective temperature. Infrared spectral
observations, however, have shown that probably some of those are not that
important, and that the whole picture is more complex.

We presented observations of young objects with a wide range in masses (or
temperatures).  In all groups, the discs are observed to show a variety in
degree of flaring, and also the dust features observed are both from
unprocessed and processed grains, as witnessed by their derived sizes and
crystallinity.

The lower the mass, the faster both the disc and dust evolution seem to
happen: for brown dwarfs most discs are flat already at an age of 5 Myr, while
the dust grains already have grown beyond a few microns at that age. Is this
because discs evolve from the inside, and in brown dwarfs we see the more 
inward regions?

For the T Tauri stars, there is a clear relation between the strength of the
10 micron feature (a measure of the size of the emitting grains) and the
spectral type: later spectral types, on average, harbour larger grains. Also
the amount of accretion, as determined by the H$\alpha$ feature, is an
indicator of the dust properties: those sources that accrete most, hence have
more turbulent discs, show the largest grain sizes in their disc atmospheres.

For the Herbig Ae/Be stars, we did not find a relation between their age and
the amount of dust processing. This could be attributed to the difficulties to
determine their ages, as they are mostly isolated.

\section{Future directions}
\label{s_future}

Obvious future steps will include a comparison of (1) the 'average' dust
properties between clusters with different ages, densities and even
metalicities, and (2) in more detail, between young objects located within
the same cluster, hence with the same initial conditions. It is only when a
meaningful statistical sample is obtained, that one can pinpoint those
properties that influence dust processing, by keeping certain variables,
e.g. environment and spectral type, as a constant. The large database of
spectra that {\it Spitzer} has provided, will certainly help to solve (pieces
of) the puzzle.

Also longer wavelength studies are important to derive the composition and
temperature of the dust, in particular for olivines at 69 micron; it is in this
context that {\it Herschel} will play an important role in the coming years.

Finally, higher spatial resolution will allow to search for the radial
dependency of dust properties within a disc, it is here that interferometers
are important. And last but not least, adaptive optics can help to resolve
binaries, so that the influence of close companions on dust and disc evolution
can be properly studied.

\acknowledgements 
Part of this work was supported by the {\it Deutsche Forschungsgemeinschaft}
(DFG) under project number ME2061/3-2.

\end{document}